\documentclass[11pt]{article}
\usepackage[T1]{fontenc}
\usepackage[utf8]{inputenc}
\usepackage{lmodern}
\usepackage[margin=1in]{geometry}
\usepackage{amsmath,amssymb}
\usepackage{graphicx}
\usepackage{float}
\usepackage{booktabs}
\usepackage{caption}
\usepackage{url}
\usepackage[hidelinks,breaklinks]{hyperref}
\usepackage{microtype}
\newsavebox{\tabbox}
\newcommand{\fittab}{\ifdim\wd\tabbox>\linewidth\resizebox{\linewidth}{!}{\usebox{\tabbox}}\else\usebox{\tabbox}\fi}

\begin{document}
\title{\textbf{Evolved Spiking Neural Networks for EEG Seizure Detection: Calcium Dynamics and Delay Mechanics}}

\author{%
  \textbf{Tuba Aksoy, PhD}\thanks{Correspondence: \href{mailto:taksoy@gmail.com}{taksoy@gmail.com}} \\
  \small Austin, TX, USA \\
  \small ORCID: \href{https://orcid.org/0000-0002-2287-6199}{0000-0002-2287-6199}
}

\date{}

\maketitle

\begin{abstract}
Neuroevolution can discover compact network designs that are hard to specify by hand. The EONS framework evolves the topology, weights, thresholds and delays of neuromorphic spiking networks without gradients. We tested whether giving integrate-and-fire (IF) neurons an activity-dependent calcium current (Ca-IF neurons) improves evolved networks for EEG seizure detection, with calcium parameters fixed, shared across a network, or evolved for each neuron. On 10 held-out CHB-MIT patients never seen in training, per-neuron calcium raised the record-level Matthews correlation coefficient (MCC) from 0.592 to 0.726 over 20 independent runs (p = 0.008) and lowered the false-alarm rate from 0.17 to 0.10. Fixed or per-network calcium gave no significant benefit. IF networks with evolved synaptic delays of 1--3 ticks gave a similar gain (p = 0.005), whereas axonal delays did not. On Siena, with unchanged settings, per-neuron calcium gained +0.096 MCC over IF in one patient-disjoint split (p = 0.011) and +0.059 in the other (p = 0.052). These results suggest that calcium and delays may make the evolutionary search more effective, and motivate testing on more data and in neuromorphic hardware.
\end{abstract}

\section{Introduction}

Neuroevolution has repeatedly shown that evolutionary search can find network designs that human engineers would not specify, and it does so without requiring gradients, which makes it applicable where backpropagation is difficult or impossible [1, ch. 11]. Evolution is particularly attractive for neuromorphic computing, where networks of spiking neurons must be compact and are intended for energy-limited devices such as wearables [1, ch. 11; 2]. Their design involves many interacting parameters (thresholds, delays, leak, spike encoding, short-term memory) that must be tuned together [1, ch. 11; 3]. The Evolutionary Optimization of Neuromorphic Systems (EONS) framework addresses this setting by evolving the topology of a spiking network together with its weights, thresholds and delays [4], and it has produced small networks that run on a low-power neuromorphic platform for time-series classification [3].

Detecting seizures in electroencephalography (EEG) is a natural target for such networks, and EONS has already been applied to it on CHB-MIT [3]. Continuous EEG monitoring generates long recordings, the events of interest are rare, and a detector that is useful in practice must keep false alarms low [5, 6]. Spiking networks have been applied to seizure detection on CHB-MIT and Siena [7, 8], but these studies evaluated them with random splits of trials or samples, so data from the same patient can appear in both training and test sets. Such evaluation is common in the field [9, 10], so performance on patients the model has never seen is less often reported. Cross-patient evaluation of spiking networks has started to appear [11], but with gradient-trained networks.

Integrate-and-fire neurons are widely used in computational neuroscience [12, 13, 14] and in neuromorphic designs, including EONS networks [1, ch. 11; 3]. In our earlier work on spiking network models of working memory and interval timing [15], adding calcium-dependent intrinsic currents to integrate-and-fire neurons made recurrent networks more robust, gave firing statistics closer to experiments, and let them learn much longer time intervals. Calcium adds a slower, activity-dependent variable (a time constant of 100 ms in our default model), which could help networks use temporal structure in EEG. We therefore examine whether such a current can also improve spiking networks designed by evolution.

We address this question for seizure detection on two public scalp-EEG cohorts, CHB-MIT and Siena, with all test patients held out from training. We compare integrate-and-fire networks with networks that have a calcium current whose parameters are fixed (Ca-IF\textsubscript{fix}), evolved per network (Ca-IF\textsubscript{net}), or evolved per neuron (Ca-IF\textsubscript{neu}). We also compare them with another route to temporal flexibility, giving the integrate-and-fire networks evolved or fixed synaptic delays as well as axonal delays. We find that calcium parameters evolved per neuron yield higher MCC than integrate-and-fire networks on CHB-MIT (20 independent runs per arm), through fewer false alarms. We then applied the same network and evolution settings to Siena, without cohort-specific tuning; the gains there are smaller (significant in one of two splits). On CHB-MIT, integrate-and-fire networks with evolved synaptic delays of 1--3 ticks gave a similar gain, while axonal delays gave no gain. Removing calcium from the evolved networks at test time did not change their scores, so the benefit may come from the evolutionary search that calcium makes possible, not from calcium dynamics in the finished networks.

\section{Methods}

Our neuroevolution pipeline mostly follows the framework of Ghawaly et al. [3], reconstructed from the published description, and adds the activity-dependent calcium current of the neuron model of Aksoy and Shouval [15] (Section 2.2). The evolved parameters are the network topology, synaptic weights and neuron thresholds, and in the calcium variants the parameters of the calcium current. Synaptic delays were not evolved in the main experiments but fixed at the minimum transmission time of one tick.

\begin{figure}[htbp]
\centering
\includegraphics[width=\linewidth]{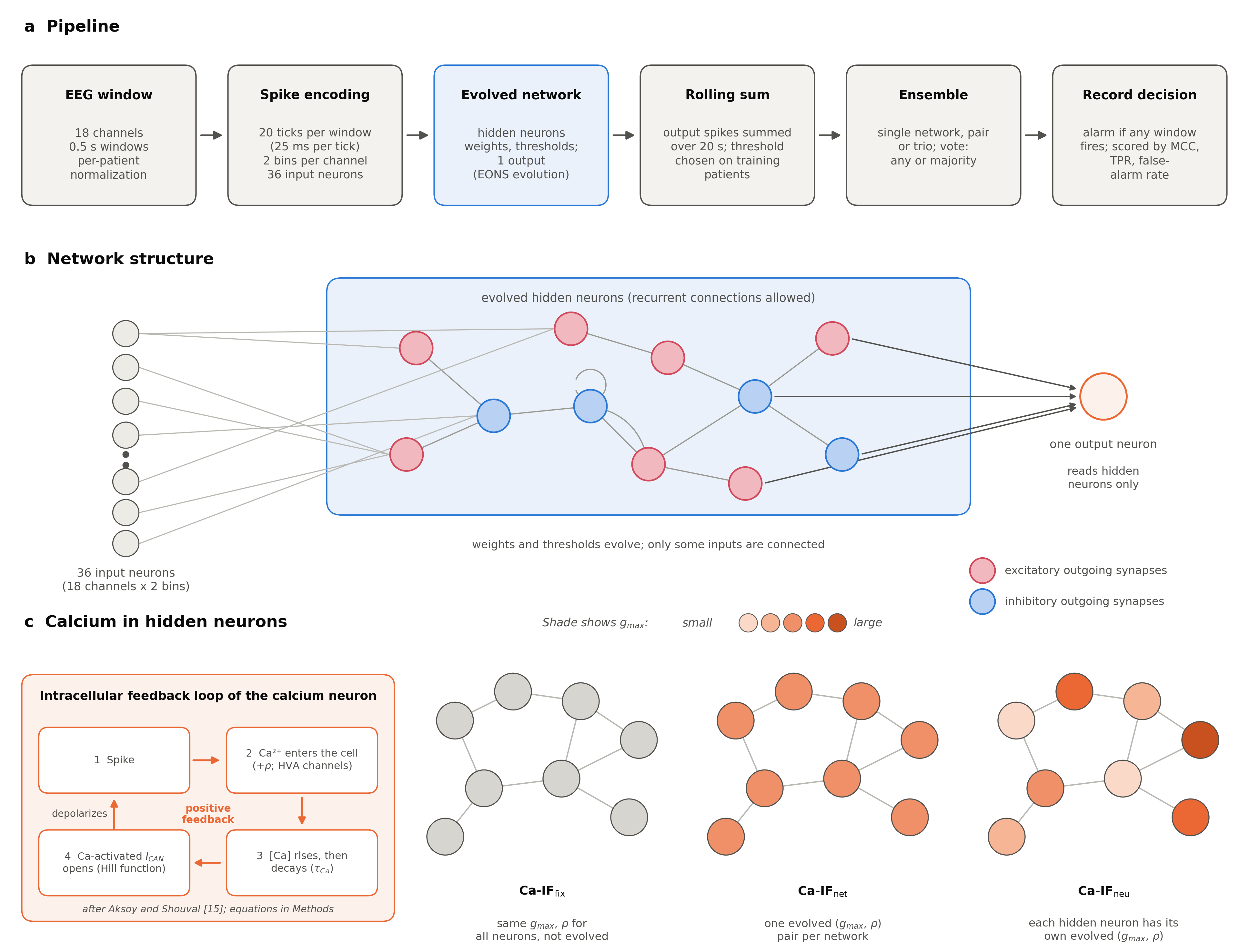}
\caption*{\textbf{Figure 1. Pipeline, network structure and calcium variants.} (a) Pipeline from EEG window to record decision; the evolved network is highlighted. (b) Network structure (schematic): input neurons connect to an evolved network of hidden neurons, recurrent connections allowed, and one output neuron reads from hidden neurons only. Colors mark the sign of the outgoing synapses (red, excitatory; blue, inhibitory). (c) The calcium feedback loop of a hidden neuron, after Aksoy and Shouval [15], and the three Ca-IF variants: Ca-IF\textsubscript{fix} (gray, fixed parameters), Ca-IF\textsubscript{net} (one evolved pair per network) and Ca-IF\textsubscript{neu} (one pair per neuron); darker shading marks a larger $g_{\max}$.}
\end{figure}

\subsection{Data Sources and Encoding Pipeline}

We used the CHB-MIT Scalp EEG database (24 cases from 23 pediatric patients aged 1.5 to 22 years, recorded at 256 Hz [16, 17]), with 18 bipolar channels. Records containing several seizures were divided into single-seizure records. Seizure records were trimmed to at most 400 s (with 8--20 s of random padding on each side of the seizure) and non-seizure records to at most 150 s. Each channel was normalized per patient using that patient's 5th and 95th percentiles. Training: chb01--05, 12, 13, 15, 16, 18. Validation: chb06, 10, 11 (not used). chb21, a later recording of the subject of chb01, was not used. Test: chb07--09, 14, 17, 19, 20, 22--24 (10 patients, 34 records, 14 with a seizure and 20 without). Ensemble selection used the training patients only.

The Siena Scalp EEG database [18, 19] contains 14 adult patients aged 20 to 71 years, recorded at 512 Hz. Both databases are available from PhysioNet [20]. We derived the same 18 bipolar channels from the referential recordings and resampled the signals to 256 Hz. Siena has no seizure-free files, so non-seizure records were cut from the longest seizure-free stretch of each file, at least 300 s from any annotated seizure, which gives 46 seizure and 41 non-seizure records (CHB-MIT test: 14 and 20). Records were then built as for CHB-MIT. We used two patient-disjoint splits. Split 1, training: PN01, PN17, PN00, PN06, PN14, PN13, PN16; validation: PN11, PN09 (not used); test: PN03, PN10, PN07, PN12, PN05 (34 records). Split 2 swaps the training and test patients of split 1 (45 test records).

Each 0.5 s window of the 18 channels was encoded as a spike train of 20 ticks (25 ms per tick), using two flip-flop bins per channel, which gives 36 input neurons (Fig. 1a, b).

\subsection{Spiking Neuron Models and Network Construction}

\textbf{Integrate-and-fire (IF).} The membrane potential accumulates weighted input spikes without leak and is bounded below at $-$1. A neuron fires and resets to 0 when its potential reaches its threshold, which is evolved between 0.1 and 1.0. Synaptic weights lie between $-$1 and 1, and each synapse has a transmission time $d$ in ticks; the minimum, $d = 1$ tick (25 ms), is used for every synapse unless stated otherwise.

\begin{equation*}
V_i(t) = \max\left\{-1,\; V_i(t-1) + \sum_j w_{ij}\, s_j(t-d_{ij})\right\} \tag{1}
\end{equation*}
\begin{equation*}
s_i(t) = \begin{cases} 1, & V_i(t) \ge \theta_i \\ 0, & \text{otherwise,} \end{cases} \qquad V_i(t) \leftarrow 0 \ \text{if}\ s_i(t) = 1
\end{equation*}

In Eq. 1, $w_{ij}$ and $d_{ij} \ge 1$ are the weight and transmission time of the synapse from neuron $j$ to neuron $i$, and $\theta_i$ is the firing threshold.

\textbf{Ca-IF neurons.} Ca-IF (calcium integrate-and-fire) neurons add the calcium-activated current of [15] to the IF neuron. Each hidden neuron has a calcium variable that decays with time constant $\tau_{Ca}$ and increases by $\rho$ after the neuron fires (Eq. 2), and a Hill function of calcium gates a current that drives the membrane potential toward the reversal level $E_{CAN}$ (Eqs. 3 and 4; Fig. 1c). We used $m = 2$, $E_{CAN} = 0.8$, $\theta_{Ca} = 2$ and $\tau_{Ca} = 4$ ticks (100 ms, as in the original model [15]). The three variants differ in how $g_{\max}$ and $\rho$ are set: the same fixed values for all neurons (Ca-IF\textsubscript{fix}; $g_{\max} = 0.05$, $\rho = 0.09$), one evolved pair per network (Ca-IF\textsubscript{net}), or one evolved pair per hidden neuron (Ca-IF\textsubscript{neu}). The values of $g_{\max}$ and $\rho$, and the upper limits of their evolved ranges, were chosen so that the calcium current does not saturate; with larger values neurons show persistent firing. Longer time constants of 16, 32, 64 and 100 ticks (0.4, 0.8, 1.6 and 2.5 s) were also tested in Ca-IF\textsubscript{neu}, with 20 runs each.

\begin{equation*}
Ca_i(t) = e^{-\Delta t/\tau_{Ca}}\, Ca_i(t-1) + \rho_i\, s_i(t-1) \tag{2}
\end{equation*}
\begin{equation*}
g_i(t) = g_{\max,i}\, \frac{Ca_i(t)^m}{Ca_i(t)^m + \theta_{Ca}^m} \tag{3}
\end{equation*}
\begin{equation*}
I_{CAN,i}(t) = -g_i(t)\,\bigl(V_i(t) - E_{CAN}\bigr), \qquad V_i(t) \leftarrow V_i(t) + I_{CAN,i}(t)\,\Delta t \tag{4}
\end{equation*}

Eqs. 2--4 are the discrete-time forms of Eqs. 10--12 of [15], with $\Delta t = 1$ tick.

\textbf{Networks.} Networks have one output neuron that reads from hidden neurons only. Recurrent connections are allowed, and each ordered pair of neurons has at most one synapse (Fig. 1b). Membrane potentials and calcium persist across the 0.5 s windows of a record, and state is reset at the start of each record.

\subsection{Evolutionary Search Setup}

A single population of $N = 100$ random networks (30 hidden neurons and 30 synapses each) was evolved for $G = 30$ generations. The fitness of a network was the Matthews correlation coefficient (MCC) of its window-level predictions on the training patients, with output threshold 0 and no rolling sum (Eq. 5):

\begin{equation*}
F(S) = \mathrm{MCC} = \frac{TP \cdot TN - FP \cdot FN}{\sqrt{(TP+FP)(TP+FN)(TN+FP)(TN+FN)}} \tag{5}
\end{equation*}

$TP$, $TN$, $FP$ and $FN$ are the numbers of true positive, true negative, false positive and false negative windows (a window is positive if the output neuron fires in it). On the test patients we report record-level MCC, the true positive rate (TPR, also called sensitivity), and the false-alarm rate (FPR), the fraction of non-seizure records that raise an alarm (Section 2.4).

Selection, crossover and mutation followed Table I of [3]: tournament selection ($k$ = 10\% of $N$, best-net factor 0.9), three elite networks, random networks with probability 0.05, crossover with probability 0.5, and four mutation operations per mutation event (adding or deleting a neuron or synapse, or perturbing a threshold or weight). In the Ca-IF variants with evolved parameters, $g_{\max}$ lay in [0, 0.07] and $\rho$ in [0, 0.20]; the values were initialized uniformly, mutated with Gaussian steps (sd 0.02), and inherited with the network or the neuron in crossover. Each independent run used a different random seed. The same network and evolution settings were used for CHB-MIT and Siena, without Siena-specific tuning.

\textbf{Delay arms.} To evaluate alternative forms of temporal flexibility, we also tested dedicated delay arms: IF networks with an evolved delay per synapse of 1--3 ticks (25--75 ms) or 1--15 ticks (25--375 ms), or a fixed delay of 4 ticks (100 ms, the default calcium time constant). In the axonal arms, one delay per neuron (as in the \textmu{}Caspian axonal delay [3, 21]) was evolved between 1 and 3 or between 1 and 15 ticks; a mutation changes the delay of one neuron. Each arm had 20 runs; the baseline used $d = 1$.

\subsection{Ensemble Selection and Performance Metrics}

At evaluation, the output spike count of each 0.5 s window, $z_t$, is summed over a trailing window of 20 s (40 windows), $r_t = \sum_{u=t-39}^{t} z_u$, truncated at the start of each record. Network $m$ predicted a seizure in window $t$ when its sum $r_t^{(m)}$ exceeded its detection threshold $\Theta_m$:

\begin{equation*}
\hat{y}_t^{(m)} = \mathbf{1}\left[\, r_t^{(m)} > \Theta_m \,\right] \tag{6}
\end{equation*}

For each network in the final population, all possible values of the detection threshold $\Theta_m$ were evaluated, and $\Theta_m$ was set to the value that maximized the window-level MCC of the rolling sum on the training patients. Networks were then ranked by that MCC and the top 10\% were retained for ensembling [3]. The best single network, pair or trio was then chosen on the training patients, with a pair voting ``either'' or ``both'' and a trio voting ``any'' or ``majority''.

The chosen ensemble was scored once on the test patients. A record was scored as alarmed, or as detected if it contained a seizure, if any window in it fired.

\subsection{Statistical Evaluation Procedures}

Every arm was run 20 times with different random seeds. Arms were compared on record-level MCC with an exact two-sided Wilcoxon signed-rank test paired by run, and we report the number of runs won, lost and tied (W--L--T). The FPR of two arms was compared across the 10 test patients with a sign test. P-values are two-sided. For each group of arms compared with IF (three calcium variants, five delay arms, four longer time constants, two Siena splits, three transfer test sets), the Bonferroni-adjusted p-value (p adj.) is also given. Ablated networks were re-scored without retraining: $g_{\max}$ was set to zero or all delays to 1, or the ($g_{\max}$, $\rho$) pairs or delays were permuted within each network.

\section{Results}

\subsection{Overview of Experiments and Core Findings}

We first compare IF networks with the Ca-IF variants on CHB-MIT (Section 3.2), then with delay and calcium time-constant variants (3.3), and test the finished networks by ablation (3.4). We then repeat the comparison on Siena (3.5) and score the CHB-MIT-trained networks on Siena (3.6). All scores are record-level MCC on held-out patients never seen during training or selection, from 20 independent runs per arm; Supplementary Table S1 lists all comparisons.

Fig. 2 shows an evolved Ca-IF\textsubscript{neu} network responding to a seizure and a non-seizure record; Supplementary Fig. S1 shows all test records of the same network.

\begin{figure}[htbp]
\centering
\includegraphics[width=0.8\linewidth]{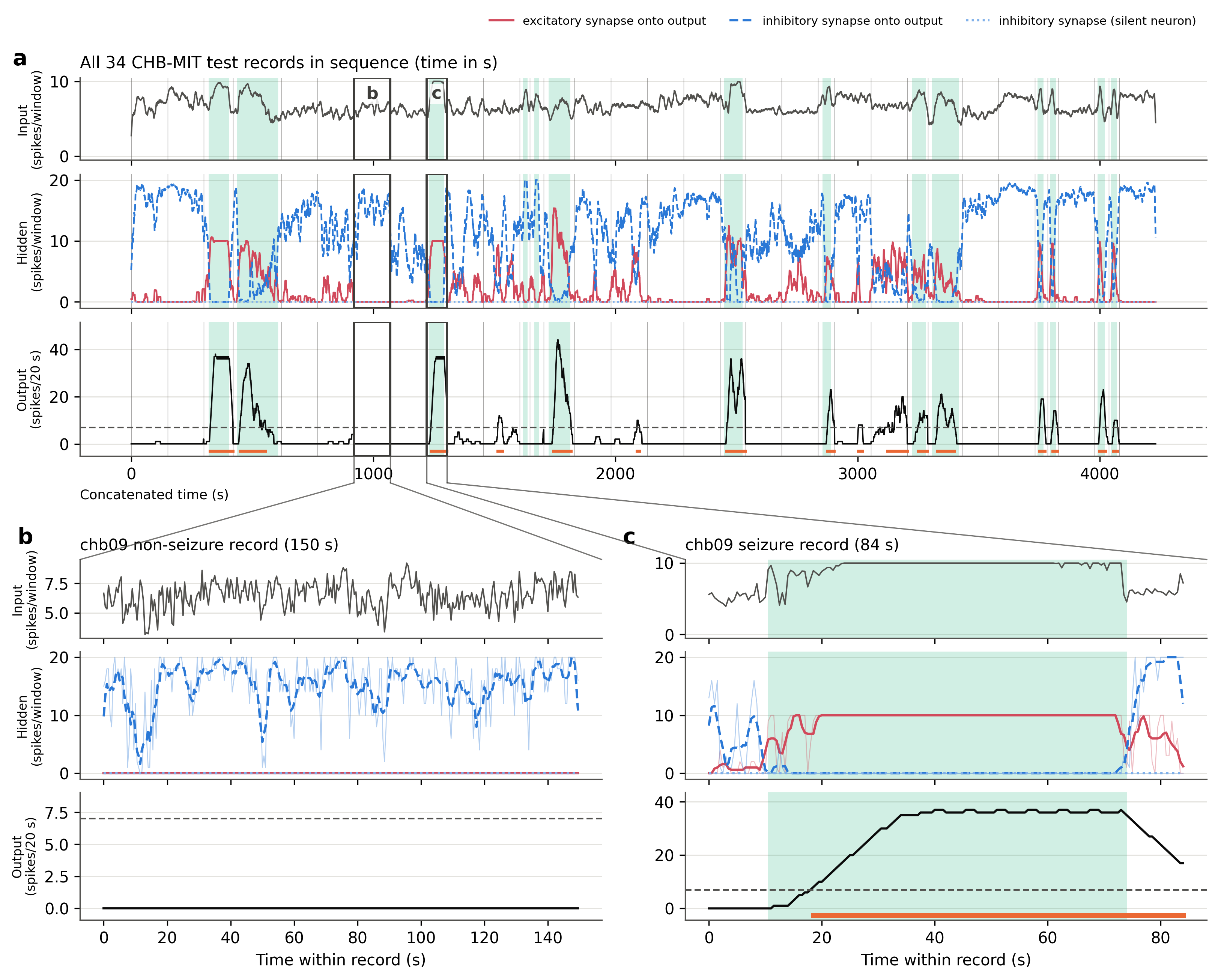}
\caption*{\textbf{Figure 2. Activity of an evolved Ca-IF\textsubscript{neu} network on CHB-MIT test records.} Run 4 (MCC 0.648). (a) All 34 test records in sequence: input firing, firing of the hidden neurons connected to the output (red, excitatory; blue, inhibitory), and the output rolling sum with the detection threshold (dashed) and alarms (orange). Green shading marks seizure windows. (b) Non-seizure record of patient chb09: no alarm. (c) Seizure record of chb09: the rolling sum crosses the threshold about 8 s after seizure onset.}
\end{figure}

\subsection{Seizure Detection Performance and False-Alarm Reduction on CHB-MIT}

On the 10 held-out CHB-MIT patients, evolving the calcium parameters per neuron (Ca-IF\textsubscript{neu}) raised MCC over IF (+0.134, p = 0.008, adjusted p = 0.025; Table 1, Fig. 3) and lowered the false-alarm rate (FPR) from 0.17 to 0.10 (Fig. 3c). Ca-IF\textsubscript{fix} and Ca-IF\textsubscript{net} gave no significant gain over IF (Fig. 3, Supplementary Table S1).

The false-alarm reduction was consistent across patients. Of the 10 held-out patients, Ca-IF\textsubscript{neu} lowered the false-alarm rate in 7, raised it in none, and left it unchanged in 3 (sign test, p = 0.016; Fig. 3d).

Network size was similar in both arms: the tested ensembles had about 24--25 synapses and 34 hidden neurons (Supplementary Table S3).

\begin{figure}[htbp]
\centering
\includegraphics[width=0.72\linewidth]{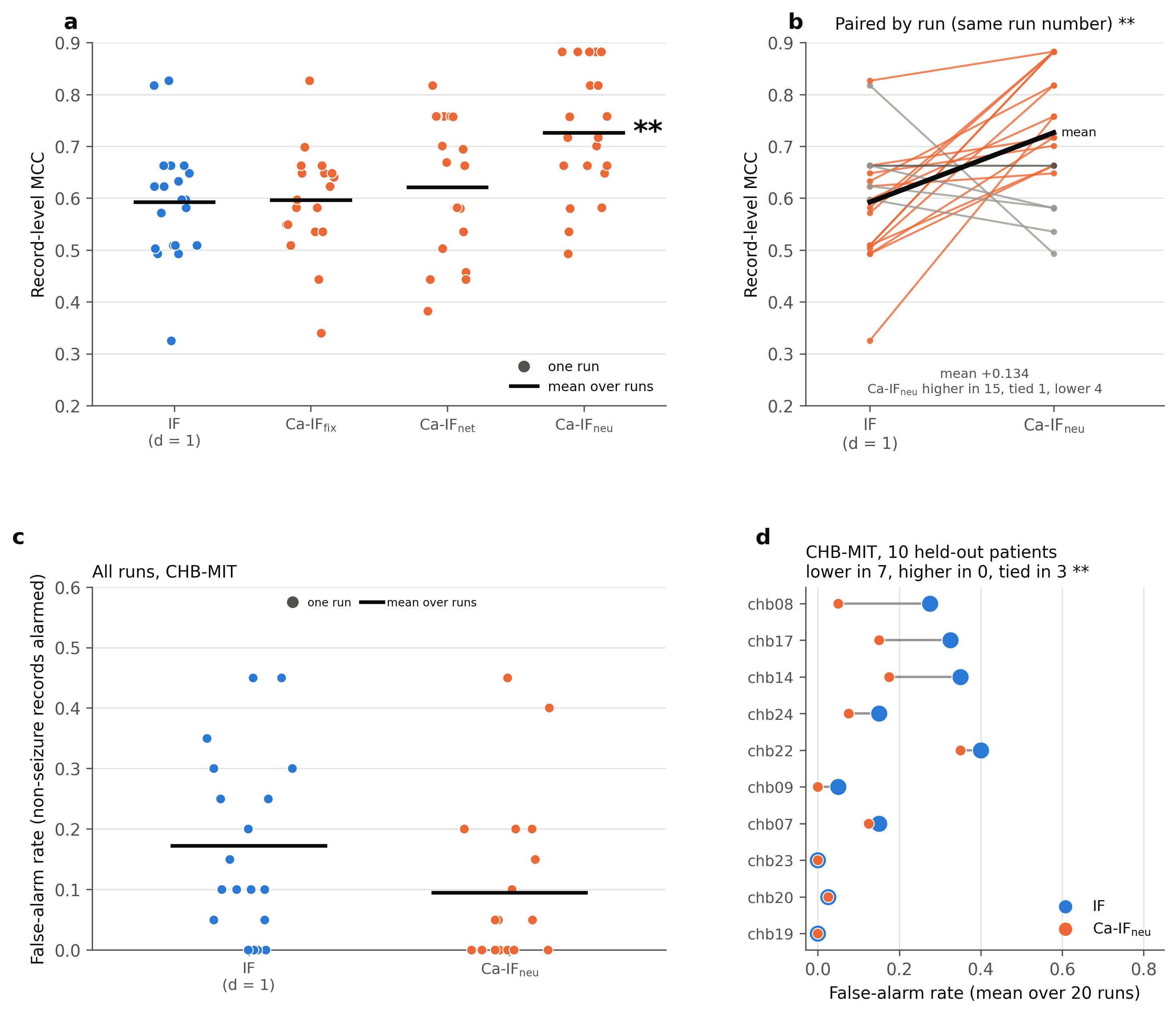}
\caption*{\textbf{Figure 3. Per-run results on the 10 held-out CHB-MIT patients.} Record-level MCC (a) for IF (d = 1), Ca-IF\textsubscript{fix}, Ca-IF\textsubscript{net} and Ca-IF\textsubscript{neu}, each dot one run, and (b) for IF against Ca-IF\textsubscript{neu}, each line joining the same run number (orange: Ca-IF\textsubscript{neu} higher, gray: lower). False-alarm rate (c) per run for IF and Ca-IF\textsubscript{neu} and (d) per held-out patient (mean over 20 runs, sorted by the difference between the arms): lower with Ca-IF\textsubscript{neu} in 7 patients, higher in none, tied in 3 (sign test p = 0.016); each patient has about two non-seizure records, hence the steps of about 0.5. Each dot is one run; bars (in b, the thick line) are the means. One star marks p $<$ 0.05 and two stars mark a Bonferroni-adjusted p $<$ 0.05 (adjusted for the three calcium variants). (a) Ca-IF\textsubscript{neu} against IF, p = 0.008, adjusted p = 0.025 (**); Ca-IF\textsubscript{fix}, p = 0.952; Ca-IF\textsubscript{net}, p = 0.430; (b) p = 0.008, adjusted p = 0.025 (**); (c) p = 0.071; (d) sign test, p = 0.016 (**). Tests across runs are paired Wilcoxon signed-rank tests.}
\end{figure}

\begin{table}[htbp]
\centering
\caption*{\textbf{Table 1. CHB-MIT: Ca-IF\textsubscript{neu} against IF (MCC difference = Ca-IF\textsubscript{neu} minus IF; 20 runs).} p adj., Bonferroni-adjusted p (Section 2.5).}
\small
\begin{lrbox}{\tabbox}
\begin{tabular}{lccccccc}
\toprule
Experiment & Test records & MCC, IF / Ca-IF\textsubscript{neu} & MCC difference & W--L--T & p & p adj. & FPR, IF / Ca-IF\textsubscript{neu} \\
\midrule
CHB-MIT & 34 & 0.592 / 0.726 & +0.134 & 15--4--1 & 0.008 & 0.025 & 0.17 / 0.10 \\
\bottomrule
\end{tabular}
\end{lrbox}\fittab
\end{table}

\subsection{Effects of Synaptic Delays and Calcium Decay Time Constants}

To test whether evolved delays reproduce the calcium gain, we compared IF networks with several delay arms (Table 2). Only evolved per-synapse delays of 1--3 ticks (25--75 ms) gave a significant gain over IF (adjusted p = 0.024), similar to that of Ca-IF\textsubscript{neu} (FPR 0.09 vs 0.10; TPR 0.77 vs 0.79). Per-synapse delays of 1--15 ticks, a fixed delay of 4 ticks, and axonal delays (one per neuron, as on \textmu{}Caspian) of 1--3 and 1--15 ticks gave smaller gains that were not significant.

\begin{table}[htbp]
\centering
\caption*{\textbf{Table 2. CHB-MIT: IF networks with different transmission times d (MCC difference = arm minus IF with d = 1; 20 runs).} p adj., Bonferroni-adjusted p (Section 2.5).}
\small
\begin{lrbox}{\tabbox}
\begin{tabular}{lccccccc}
\toprule
Arm & MCC & TPR & FPR & MCC difference & W--L--T & p & p adj. \\
\midrule
IF (d = 1) & 0.592 $\pm$ 0.111 & 0.74 & 0.17 &  &  &  &  \\
Ca-IF\textsubscript{neu} (for comparison) & 0.726 $\pm$ 0.121 & 0.79 & 0.10 & +0.134 & 15--4--1 & 0.008 & 0.025 \\
IF, evolved d 1--3 ticks, per synapse & 0.703 $\pm$ 0.116 & 0.77 & 0.09 & +0.111 & 15--3--2 & 0.005 & 0.024 \\
IF, evolved d 1--15 ticks, per synapse & 0.636 $\pm$ 0.126 & 0.75 & 0.14 & +0.044 & 12--8--0 & 0.409 & 1.000 \\
IF, fixed d = 4 ticks, per synapse & 0.658 $\pm$ 0.135 & 0.78 & 0.14 & +0.065 & 14--3--3 & 0.089 & 0.443 \\
IF, evolved d 1--3 ticks, per axon & 0.629 $\pm$ 0.133 & 0.76 & 0.15 & +0.036 & 9--10--1 & 0.515 & 1.000 \\
IF, evolved d 1--15 ticks, per axon & 0.640 $\pm$ 0.137 & 0.78 & 0.15 & +0.048 & 12--8--0 & 0.498 & 1.000 \\
\bottomrule
\end{tabular}
\end{lrbox}\fittab
\end{table}

To test the effect of the calcium time constant, we evolved Ca-IF\textsubscript{neu} with longer time constants than the default of 4 ticks. With 16 ticks (400 ms) the gain over IF was smaller than with the default and significant only before correction (p = 0.038, adjusted p = 0.15); with 32, 64 and 100 ticks it was smaller and not significant (Supplementary Table S1, Fig. 4).

\begin{table}[htbp]
\centering
\includegraphics[width=0.6\linewidth]{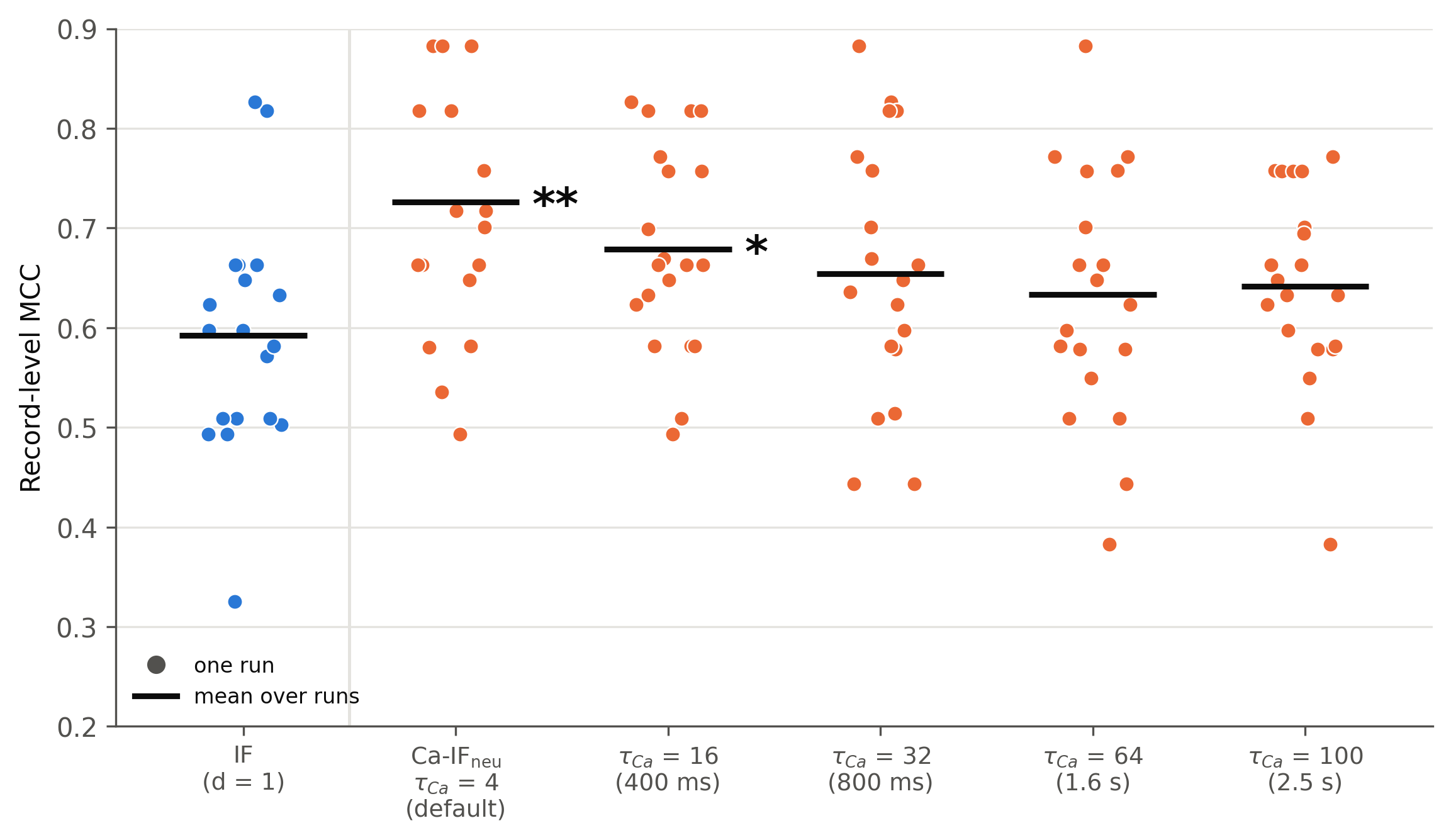}
\caption*{\textbf{Figure 4. Record-level MCC for different calcium time constants.} IF (d = 1) and Ca-IF\textsubscript{neu} with $\tau_{Ca}$ of 4 ticks (default) and 16, 32, 64 or 100 ticks; each dot is one run (20 runs) and the bars are the means. One star marks p $<$ 0.05 and two stars a Bonferroni-adjusted p $<$ 0.05, against IF (paired by run): 4 ticks, p = 0.008, adjusted p = 0.025 (**; adjusted for the three calcium variants); 16 ticks, p = 0.038, adjusted p = 0.15 (*); 32 ticks, p = 0.119, adjusted p = 0.48; 64 ticks, p = 0.304, adjusted p = 1.0; 100 ticks, p = 0.169, adjusted p = 0.68 (adjusted for the four longer time constants).}
\end{table}

\subsection{Parameter Ablation and Post-Training Network Reliance}

To test whether the finished networks depend on their calcium and their evolved delays, we changed these parameters at test time, without retraining. Removing the calcium current from the Ca-IF\textsubscript{neu} networks, or shuffling the calcium parameters among their neurons, did not change MCC (identical in all 20 runs); nor did setting the evolved delays of the IF networks to 1, or shuffling them (changes below 0.01; Supplementary Table S2). With calcium or delays removed, both kinds of network still scored higher than the IF baseline (d = 1).

\subsection{Within-Cohort Validation on the Siena Database}

To check whether the gain holds on a second cohort, we repeated the experiment on Siena with the same network and evolution settings as for CHB-MIT. Networks were trained on some Siena patients and tested on others, in two different splits (Section 2.1, Table 3, Fig. 5). Ca-IF\textsubscript{neu} gained +0.059 over IF in split 1 (p = 0.052, adjusted p = 0.105) and +0.096 in split 2 (p = 0.011, adjusted p = 0.021).

In split 1, TPR was similar (0.92 vs 0.89) and the false-alarm rate fell from 0.74 to 0.65, but it remained high. In split 2, the false-alarm rate fell from 0.39 to 0.33 and TPR rose from 0.75 to 0.79 (Fig. 5c), higher in 3 of the 7 test patients and lower in 1.

\begin{figure}[H]
\centering
\includegraphics[width=\linewidth]{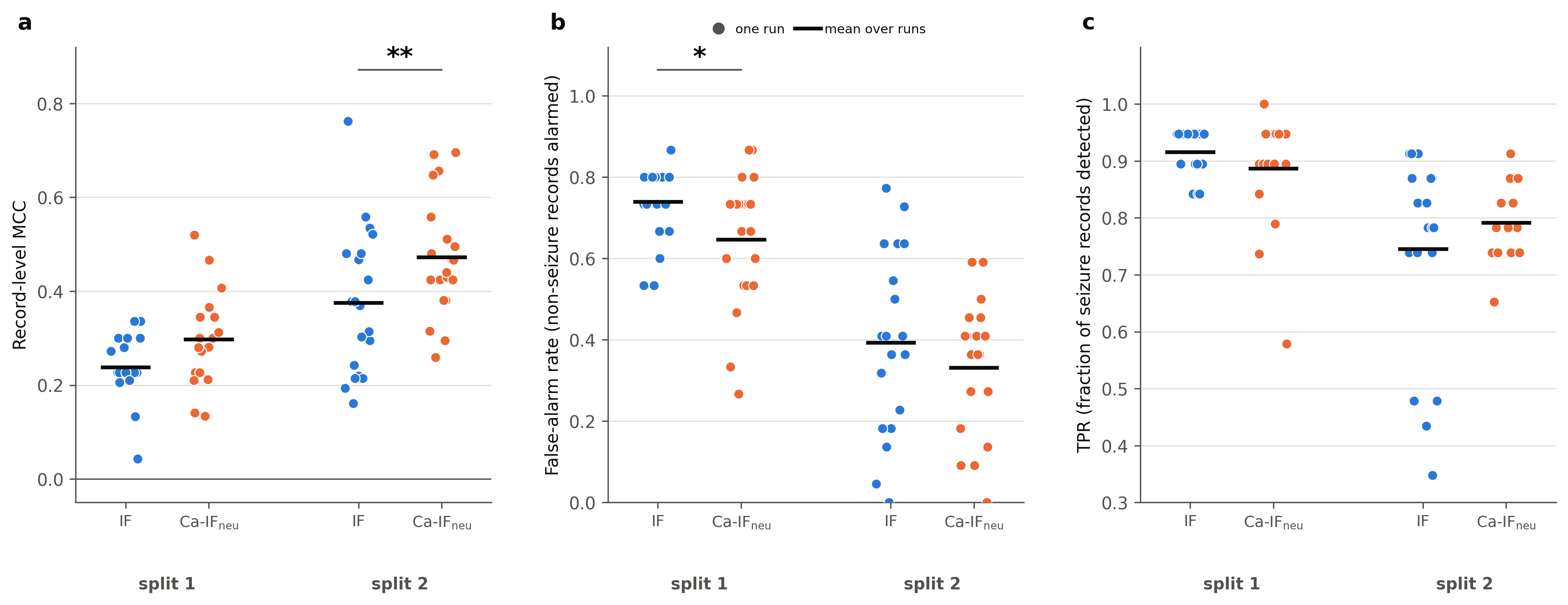}
\caption*{\textbf{Figure 5. Siena: networks trained and tested on Siena patients.} (a) Record-level MCC, (b) false-alarm rate and (c) TPR per run for IF and Ca-IF\textsubscript{neu} in split 1 (34 test records) and split 2 (45 test records). Each dot is one run; bars are the means. One star marks p $<$ 0.05 and two stars a Bonferroni-adjusted p $<$ 0.05 (two splits) for Ca-IF\textsubscript{neu} against IF (paired by run): MCC, split 1 p = 0.052 (adjusted 0.105), split 2 p = 0.011 (adjusted 0.021, **); FPR, split 1 p = 0.048 (adjusted 0.096, *), split 2 p = 0.241 (adjusted 0.48); TPR, split 1 p = 0.365 (adjusted 0.73), split 2 p = 0.713 (adjusted 1.0).}
\end{figure}

\begin{table}[htbp]
\centering
\caption*{\textbf{Table 3. Siena, networks trained and tested on Siena patients (two patient splits; 20 runs; MCC difference = Ca-IF\textsubscript{neu} minus IF).} p adj., Bonferroni-adjusted p (Section 2.5).}
\small
\begin{lrbox}{\tabbox}
\begin{tabular}{llccccccc}
\toprule
Trained on & Test set & Test records & MCC, IF / Ca-IF\textsubscript{neu} & MCC difference & W-L-T & p & p adj. & FPR, IF / Ca-IF\textsubscript{neu} \\
\midrule
Siena split 1 training patients & Siena split 1 test patients & 34 & 0.238 / 0.297 & +0.059 & 12--6--2 & 0.052 & 0.105 & 0.74 / 0.65 \\
Siena split 2 training patients & Siena split 2 test patients & 45 & 0.376 / 0.472 & +0.096 & 15--5--0 & 0.011 & 0.021 & 0.39 / 0.33 \\
\bottomrule
\end{tabular}
\end{lrbox}\fittab
\end{table}

\subsection{Cross-Cohort Generalization and Direct Transfer to Siena}

To test transfer between cohorts, we scored the CHB-MIT-trained ensembles (20 runs) on all 14 Siena patients (87 records) without retraining. They reached MCC 0.416 (IF) and 0.469 (Ca-IF\textsubscript{neu}), lower than on CHB-MIT, and the Ca-IF\textsubscript{neu} gain (+0.053) was not significant (p = 0.087; Table 4). On the Siena test patients of splits 1 and 2, however, these ensembles scored similarly to or higher than networks trained on Siena itself (Table 4, Table 3).

\begin{table}[htbp]
\centering
\caption*{\textbf{Table 4. CHB-MIT-trained ensembles scored on Siena without retraining (20 runs; MCC difference = Ca-IF\textsubscript{neu} minus IF).} p adj., Bonferroni-adjusted p (Section 2.5).}
\small
\begin{lrbox}{\tabbox}
\begin{tabular}{llccccccc}
\toprule
Trained on & Test set & Test records & MCC, IF / Ca-IF\textsubscript{neu} & MCC difference & W-L-T & p & p adj. & FPR, IF / Ca-IF\textsubscript{neu} \\
\midrule
CHB-MIT & All 14 Siena patients & 87 & 0.416 / 0.469 & +0.053 & 12--7--1 & 0.087 & 0.262 & 0.29 / 0.23 \\
CHB-MIT & Siena split 1 test patients & 34 & 0.261 / 0.336 & +0.075 & 15--5--0 & 0.076 & 0.228 & 0.36 / 0.31 \\
CHB-MIT & Siena split 2 test patients & 45 & 0.533 / 0.561 & +0.029 & 11--8--1 & 0.374 & 1.000 & 0.23 / 0.17 \\
\bottomrule
\end{tabular}
\end{lrbox}\fittab
\end{table}

\section{Discussion}

We tested integrate-and-fire networks for seizure detection with calcium parameters that were fixed, shared across a network, or evolved per neuron (Ca-IF\textsubscript{neu}), and with evolved delays. The test patients were CHB-MIT patients held out from training, validation and selection. Calcium parameters evolved per neuron gave higher MCC and fewer false alarms than integrate-and-fire networks, consistently across patients (Fig. 3d). Short evolved synaptic delays gave a similar gain, whereas longer or fixed delays and axonal delays gave no significant gain.

In biological neurons, calcium entry and the currents it gates differ from neuron to neuron and adapt with activity [15, 22]. Bringing this variability into the evolutionary search, through per-neuron calcium and synaptic delays, improved on the baseline, whereas calcium parameters that were fixed or shared across a network did not (Supplementary Table S1). This also agrees with neuroevolution findings that evolution can tune many aspects of a design at once [1, ch. 11], and that letting components such as activation functions differ across parts of a network improves performance [23]. Among the time constants tested, the default of 100 ms, taken from the biophysical model of Aksoy and Shouval [15], gave the largest gain (p = 0.008).

Removing or shuffling the calcium parameters of Ca-IFneu networks, or the evolved synaptic delays of IF networks, at test time did not change their performance appreciably. One possible explanation is that these parameters add heritable variation that helps the evolutionary search reach better networks (topology, weights, thresholds), even though they are not needed afterwards. Testing this would require focused experiments, which we leave for future work. If this holds, evolution in software could tune these parameters for each network, and possibly for each patient, and the finished networks could run in hardware without a calcium current or synaptic delays.

On Siena, we trained networks using the same network and evolution settings as for CHB-MIT, without Siena-specific tuning. The MCC gain was smaller than on CHB-MIT. The smaller gain may reflect patient age (adults versus pediatric patients), the recording setup, the number of patients, or the absence of cohort-specific tuning.

We also scored the CHB-MIT-trained networks on Siena without retraining (Section 3.6, Table 4). Performance dropped relative to CHB-MIT, as has been reported for other seizure detectors tested on a different dataset [24], and the Ca-IFneu advantage over IF was not significant. On both Siena splits, these networks scored similarly to or higher than networks trained on Siena itself, which had 5 to 7 training patients against 10 for CHB-MIT. A larger training cohort may therefore help networks transfer to an unseen dataset, although demographics and recording settings may also affect transfer.

A primary limitation of this study is the modest cohort size and the limited number of seizure records. Each split used one training set, so the effect of the choice of training patients was not examined. All experiments were conducted in software. We report record-level metrics; event-level measures (detection latency, false alarms per hour) and patient-level cross-validation were not evaluated. Further work is required to address these limitations and to make this approach applicable to neuromorphic hardware. While calcium evolved per neuron gave consistent gains on held-out CHB-MIT data, confirming its broader clinical utility will require evaluation on larger, real-world patient populations.

\section{Conclusion}

Evolving calcium parameters per neuron in integrate-and-fire networks gave seizure detectors with higher MCC and fewer false alarms on held-out CHB-MIT patients, and smaller gains on Siena with unchanged settings (significant in one of two splits). Evolved synaptic delays of 1--3 ticks gave a similar gain on CHB-MIT, while axonal delays gave no gain. Both calcium and evolved delays may act by helping evolution find better networks, not by changing how the finished networks work. These software results on small cohorts motivate larger evaluations and, next, neuromorphic hardware.

\section*{Declarations}

\textbf{Data availability.} The CHB-MIT Scalp EEG Database [16, 17] and the Siena Scalp EEG Database [18, 19] are publicly available from PhysioNet [20]. The spike-encoded records and the evolved networks were generated with the code described below.

\textbf{Code availability.} Code is available from the author on reasonable request.

\textbf{Funding.} This work received no external funding.

\textbf{Competing interests.} The author declares no competing interests.

\clearpage
\section*{Supplementary tables and figures}

Figure S1 uses the CHB-MIT-trained Ca-IF\textsubscript{neu} ensemble of run 4 (record-level MCC 0.648), the network shown in Figure 2; its chosen ensemble is a single network with detection threshold 7.

\begin{table}[htbp]
\centering
\caption*{\textbf{Supplementary Table S1. All results, 20 runs per arm. MCC difference, W--L--T and p are against the IF baseline of the same test set (IF, d = 1).}}
\small
\begin{lrbox}{\tabbox}
\begin{tabular}{lccccccc}
\toprule
Arm & MCC & TPR & FPR & MCC difference & W--L--T & p & p adj. \\
\midrule
\textbf{\textbf{CHB-MIT (34 test records)}} \\
IF (d = 1) & 0.592 $\pm$ 0.111 & 0.74 & 0.17 &  &  &  &  \\
Ca-IF\textsubscript{fix} & 0.596 $\pm$ 0.099 & 0.75 & 0.17 & +0.004 & 7--7--6 & 0.952 & 1.000 \\
Ca-IF\textsubscript{net} & 0.621 $\pm$ 0.127 & 0.74 & 0.14 & +0.029 & 11--9--0 & 0.430 & 1.000 \\
Ca-IF\textsubscript{neu}, $\tau_{Ca}$ = 4 ticks (default) & 0.726 $\pm$ 0.121 & 0.79 & 0.10 & +0.134 & 15--4--1 & 0.008 & 0.025 \\
Ca-IF\textsubscript{neu}, $\tau_{Ca}$ = 16 ticks & 0.679 $\pm$ 0.100 & 0.74 & 0.10 & +0.086 & 13--5--2 & 0.038 & 0.154 \\
Ca-IF\textsubscript{neu}, $\tau_{Ca}$ = 32 ticks & 0.654 $\pm$ 0.125 & 0.77 & 0.14 & +0.061 & 12--6--2 & 0.119 & 0.475 \\
Ca-IF\textsubscript{neu}, $\tau_{Ca}$ = 64 ticks & 0.633 $\pm$ 0.120 & 0.75 & 0.14 & +0.041 & 10--8--2 & 0.304 & 1.000 \\
Ca-IF\textsubscript{neu}, $\tau_{Ca}$ = 100 ticks & 0.642 $\pm$ 0.096 & 0.77 & 0.14 & +0.049 & 13--6--1 & 0.169 & 0.675 \\
IF, evolved d 1--3 ticks, per synapse & 0.703 $\pm$ 0.116 & 0.77 & 0.09 & +0.111 & 15--3--2 & 0.005 & 0.024 \\
IF, evolved d 1--15 ticks, per synapse & 0.636 $\pm$ 0.126 & 0.75 & 0.14 & +0.044 & 12--8--0 & 0.409 & 1.000 \\
IF, fixed d = 4 ticks, per synapse & 0.658 $\pm$ 0.135 & 0.78 & 0.14 & +0.065 & 14--3--3 & 0.089 & 0.443 \\
IF, evolved d 1--3 ticks, per axon & 0.629 $\pm$ 0.133 & 0.76 & 0.15 & +0.036 & 9--10--1 & 0.515 & 1.000 \\
IF, evolved d 1--15 ticks, per axon & 0.640 $\pm$ 0.137 & 0.78 & 0.15 & +0.048 & 12--8--0 & 0.498 & 1.000 \\
\textbf{\textbf{CHB-MIT, ablations of trained networks}} \\
Ca-IF\textsubscript{neu}, intact & 0.726 $\pm$ 0.121 & 0.79 & 0.10 &  &  &  \\
Ca-IF\textsubscript{neu}, calcium off & 0.726 $\pm$ 0.121 & 0.79 & 0.10 &  &  &  \\
Ca-IF\textsubscript{neu}, calcium shuffled (mean of 5) & 0.726 $\pm$ 0.121 & 0.79 & 0.10 &  &  &  \\
IF, evolved d 1--3, intact & 0.703 $\pm$ 0.116 & 0.77 & 0.09 &  &  &  \\
IF, evolved d 1--3, all delays set to 1 & 0.696 $\pm$ 0.113 & 0.78 & 0.10 &  &  &  \\
IF, evolved d 1--3, delays shuffled (mean of 5) & 0.705 $\pm$ 0.120 & 0.77 & 0.09 &  &  &  \\
\textbf{\textbf{Siena-trained, split 1 (34 records)}} \\
IF & 0.238 $\pm$ 0.065 & 0.92 & 0.74 &  &  &  &  \\
Ca-IF\textsubscript{neu} & 0.297 $\pm$ 0.094 & 0.89 & 0.65 & +0.059 & 12--6--2 & 0.052 & 0.105 \\
\textbf{\textbf{Siena-trained, split 2 (45 records)}} \\
IF & 0.376 $\pm$ 0.150 & 0.75 & 0.39 &  &  &  &  \\
Ca-IF\textsubscript{neu} & 0.472 $\pm$ 0.122 & 0.79 & 0.33 & +0.096 & 15--5--0 & 0.011 & 0.021 \\
\textbf{\textbf{CHB-MIT-trained, Siena all (87 records)}} \\
IF & 0.416 $\pm$ 0.085 & 0.71 & 0.29 &  &  &  &  \\
Ca-IF\textsubscript{neu} & 0.469 $\pm$ 0.069 & 0.69 & 0.23 & +0.053 & 12--7--1 & 0.087 & 0.262 \\
\textbf{\textbf{CHB-MIT-trained, Siena split 1 (34)}} \\
IF & 0.261 $\pm$ 0.150 & 0.61 & 0.36 &  &  &  &  \\
Ca-IF\textsubscript{neu} & 0.336 $\pm$ 0.104 & 0.64 & 0.31 & +0.075 & 15--5--0 & 0.076 & 0.228 \\
\textbf{\textbf{CHB-MIT-trained, Siena split 2 (45)}} \\
IF & 0.533 $\pm$ 0.093 & 0.76 & 0.23 &  &  &  &  \\
Ca-IF\textsubscript{neu} & 0.561 $\pm$ 0.088 & 0.72 & 0.17 & +0.029 & 11--8--1 & 0.374 & 1.000 \\
\bottomrule
\end{tabular}
\end{lrbox}\fittab
\par\smallskip
\begin{minipage}{\linewidth}\footnotesize MCC is the mean $\pm$ SD over runs (divisor n). TPR and FPR are pooled over the runs' records (CHB-MIT 34 test records: 14 with a seizure and 20 non-seizure). Siena-trained: networks trained and tested on Siena patients; CHB-MIT-trained: ensembles scored on Siena without retraining (split 1 and split 2 test patients, number of records in brackets). Exact two-sided Wilcoxon signed-rank test paired by run. p adj., Bonferroni-adjusted p within each group of arms compared with IF (three calcium variants, five delay arms, four longer time constants, two Siena splits, three transfer test sets).\end{minipage}
\end{table}

\begin{table}[htbp]
\centering
\caption*{\textbf{Supplementary Table S2. CHB-MIT: ablation of calcium and of evolved delays in the trained networks (20 runs).}}
\small
\begin{lrbox}{\tabbox}
\begin{tabular}{llccc}
\toprule
Networks & Condition & MCC & TPR & FPR \\
\midrule
IF, d = 1 & baseline & 0.592 $\pm$ 0.111 & 0.74 & 0.17 \\
Ca-IF\textsubscript{neu} & intact & 0.726 $\pm$ 0.121 & 0.79 & 0.10 \\
Ca-IF\textsubscript{neu} & calcium off & 0.726 $\pm$ 0.121 & 0.79 & 0.10 \\
Ca-IF\textsubscript{neu} & calcium shuffled (mean of 5) & 0.726 $\pm$ 0.121 & 0.79 & 0.10 \\
IF, evolved d 1--3 & intact & 0.703 $\pm$ 0.116 & 0.77 & 0.09 \\
IF, evolved d 1--3 & all delays set to 1 & 0.696 $\pm$ 0.113 & 0.78 & 0.10 \\
IF, evolved d 1--3 & delays shuffled (mean of 5) & 0.705 $\pm$ 0.120 & 0.77 & 0.09 \\
\bottomrule
\end{tabular}
\end{lrbox}\fittab
\par\smallskip
\begin{minipage}{\linewidth}\footnotesize Trained networks were re-scored without retraining. MCC is the mean $\pm$ SD over runs (divisor n). TPR and FPR are pooled over the runs' records (CHB-MIT test set, 34 records).\end{minipage}
\end{table}

\begin{table}[htbp]
\centering
\caption*{\textbf{Supplementary Table S3. CHB-MIT: size of the tested ensembles, IF (d = 1) and Ca-IF\textsubscript{neu} (mean over the networks of each run's tested ensemble).}}
\small
\begin{lrbox}{\tabbox}
\begin{tabular}{lcccc}
\toprule
Statistic & IF / Ca-IF\textsubscript{neu} & Difference & W--L--T & p \\
\midrule
Hidden neurons & 33.6 / 34.1 & +0.5 & 11--9--0 & 0.805 \\
Synapses & 24.0 / 25.3 & +1.3 & 10--10--0 & 0.721 \\
Input$\rightarrow$hidden synapses & 14.5 / 15.0 & +0.5 & 10--9--1 & 0.806 \\
Hidden$\rightarrow$hidden synapses & 7.1 / 8.0 & +0.8 & 12--8--0 & 0.447 \\
Hidden$\rightarrow$output synapses & 2.08 / 1.98 & $-$0.10 & 5--7--8 & 0.771 \\
Self-loops & 0.20 / 0.49 & +0.29 & 7--1--12 & 0.062 \\
Mean hidden threshold & 0.557 / 0.568 & +0.011 & 13--7--0 & 0.261 \\
\bottomrule
\end{tabular}
\end{lrbox}\fittab
\par\smallskip
\begin{minipage}{\linewidth}\footnotesize Ensembles have 1.9 networks on average in both arms. Difference = Ca-IF\textsubscript{neu} minus IF; exact two-sided Wilcoxon signed-rank test paired by run (20 runs).\end{minipage}
\end{table}

\clearpage
\begin{figure}[p]
\centering
\includegraphics[width=\linewidth,height=0.78\textheight,keepaspectratio]{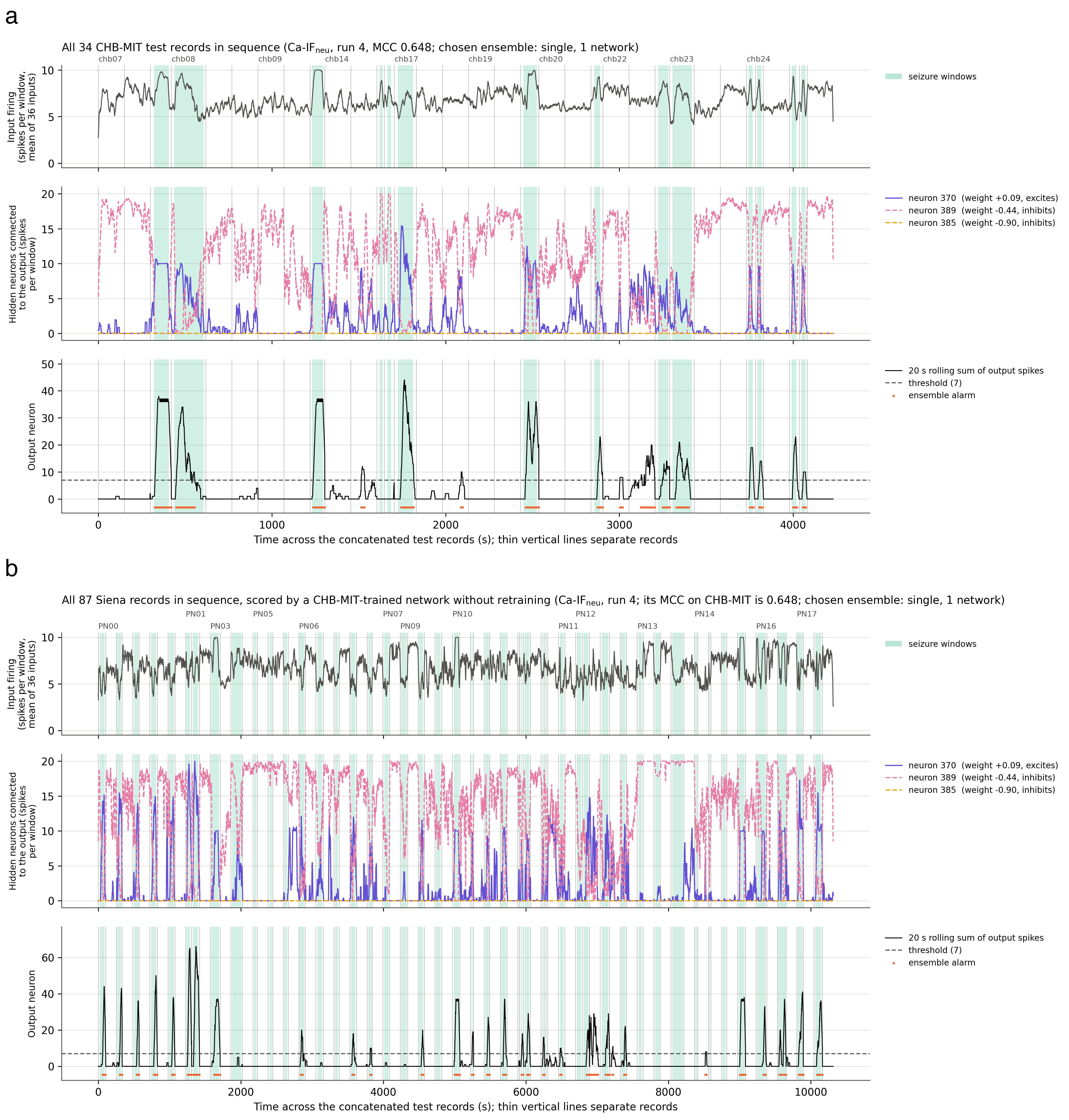}
\caption*{\textbf{Figure S1. The CHB-MIT-trained network of Figure 2 (Ca-IF\textsubscript{neu}, run 4) on all CHB-MIT test records and all Siena records.} (a) All 34 CHB-MIT test records and (b) all 87 Siena records (14 patients), scored without retraining. Rows as in Figure 2a: input firing, hidden neurons connected to the output, and the output rolling sum with the detection threshold (7) and alarms. Green shading marks seizure windows; in b, patient names are staggered above the top panel.}
\end{figure}

\end{document}